\documentclass[aps,pre,twocolumn,showpacs,preprintnumbers,amsmath,amssymb,superscriptaddress]{revtex4}
\usepackage{dcolumn}                    
\usepackage{bm}                        
\usepackage{graphicx}
\usepackage{epstopdf}
\begin{document}

\newcommand{\Imag}{{\Im\mathrm{m}}}   
\newcommand{\Real}{{\mathrm{Re}}}   
\newcommand{\im}{\mathrm{i}}        
\newcommand{\talpha}{\tilde{\alpha}}
\newcommand{\ve}[1]{{\mathbf{#1}}}

\newcommand{\x}{\lambda}  
\newcommand{\y}{\rho}     
\newcommand{\T}{\mathrm{T}}   
\newcommand{\Pv}{\mathcal{P}} 
\newcommand{\vk}{\ve{k}} 
\newcommand{\vp}{\ve{p}} 

\newcommand{\N}{\underline{\mathcal{N}}} 
\newcommand{\Nt}{\underline{\tilde{\mathcal{N}}}} 
\newcommand{\g}{\underline{\gamma}} 
\newcommand{\gt}{\underline{\tilde{\gamma}}} 

\newcommand{\vecr}{\ve{r}} 
\newcommand{\vq}{\ve{q}} 
\newcommand{\ca}[2][]{c_{#2}^{\vphantom{\dagger}#1}} 
\newcommand{\cc}[2][]{c_{#2}^{{\dagger}#1}}          
\newcommand{\da}[2][]{d_{#2}^{\vphantom{\dagger}#1}} 
\newcommand{\dc}[2][]{d_{#2}^{{\dagger}#1}}          
\newcommand{\ga}[2][]{\gamma_{#2}^{\vphantom{\dagger}#1}} 
\newcommand{\gc}[2][]{\gamma_{#2}^{{\dagger}#1}}          
\newcommand{\ea}[2][]{\eta_{#2}^{\vphantom{\dagger}#1}} 
\newcommand{\ec}[2][]{\eta_{#2}^{{\dagger}#1}}          
\newcommand{\su}{\uparrow}    
\newcommand{\sd}{\downarrow}  
\newcommand{\Tkp}[1]{T_{\vk\vp#1}}  
\newcommand{\muone}{\mu^{(1)}}      
\newcommand{\mutwo}{\mu^{(2)}}      
\newcommand{\epsk}{\varepsilon_\vk}
\newcommand{\epsp}{\varepsilon_\vp}
\newcommand{\e}[1]{\mathrm{e}^{#1}}
\newcommand{\dif}{\mathrm{d}} 
\newcommand{\diff}[2]{\frac{\dif #1}{\dif #2}}
\newcommand{\pdiff}[2]{\frac{\partial #1}{\partial #2}}
\newcommand{\mean}[1]{\langle#1\rangle}
\newcommand{\abs}[1]{|#1|}
\newcommand{\abss}[1]{|#1|^2}
\newcommand{\Sk}[1][\vk]{\ve{S}_{#1}}
\newcommand{\pauli}[1][\alpha\beta]{\boldsymbol{\sigma}_{#1}^{\vphantom{\dagger}}}

\newcommand{\eq}{Eq.}
\newcommand{\eqs}{Eqs.}
\newcommand{\cf}{\textit{cf. }}
\newcommand{\ie}{\textit{i.e. }}
\newcommand{\eg}{\textit{e.g. }}
\newcommand{\etal}{\emph{et al.}}
\def\i{\mathrm{i}}

\title{Chirality Sensitive Domain Wall Motion in Spin-Orbit Coupled Ferromagnets}

\author{Jacob Linder}
\affiliation{Department of Physics, Norwegian University of
Science and Technology, N-7491 Trondheim, Norway}

\date{\today}

\begin{abstract}
Using the Lagrangian formalism, we solve analytically the equations of motion for current-induced domain-wall dynamics in a ferromagnet with Rashba spin-orbit coupling. An exact solution for the domain wall velocity is provided, including the effect of non-equilibrium conduction electron spin-density, Gilbert damping, and the Rashba interaction parameter. We demonstrate explicitly that the influence of spin-orbit interaction can be qualitatively different from the role of non-adiabatic spin-torque in the sense that the former is sensitive to the chirality of the domain wall whereas the latter is not: the domain wall velocity shows a reentrant behavior upon changing the chirality of the domain wall. This could be used to experimentally distinguish between the spin-orbit and non-adiabatic contribution to the wall speed. A quantitative estimate for the attainable domain wall velocity is given, based on an experimentally relevant set of parameters for the system.

\end{abstract}
\maketitle

\section{Introduction}

The study of domain-wall motion in ferromagnetic materials has attracted much interest in recent years. Besides its allure from a fundamental physics viewpoint, electric control of magnetic textures is attractive in terms of potential new applications such as magnetic memory. A key concept in the context of domain wall motion is the so-called spin-transfer torque \cite{slon, berger_prb_96, stiles_prb_02}: in essence, it consists of a transfer of a transverse spin-current component to the ferromagnetic order parameter which may occur in a non-collinear magnetization configuration. Active control over domain-wall motion is a chief objective in terms of realizing the "magnetic race-track" technology put forward in \cite{racetrack}. Other ways to manipulate domain wall motion include energy redistribution in the presence of an external magnetic field, by applying microwave radiation \cite{schryer_jap_74, atkinson_nature_03, beach_prl_06, wang_ann_09, yan_prb_09}, and by using magnons \cite{han_apl_09, seo_apl_11, yan_prl_11, linder_prb_12}. Moreover, the concept of spin-transfer torque has recently been studied in antiferromagnets \cite{macdonald, haney_prb_07, xu_prl_08, gomonay_prb, linder_prb_11, hals_prl_11} including the possibility to move domain-walls.

Interestingly, it turns out that spin-orbit interactions can substantially modify both the spin-transfer torque and the resulting domain wall motion \cite{manchon_prb, obata_prb_08, ryu_jmm_12}. The combined influence of a magnetic exchange field together with spin-orbit interaction, typically taken in the Rashba form, can be shown to give rise to a non-equilibrium spin density perpendicular to the injected current flow. In turn, this gives rise to an effective magnetic field which causes magnetization dynamics and, for sufficiently strong current density, magnetization switching. The effective "spin-orbit torque" arising in this manner is qualitatively different from the conventional spin-transfer torque due to the different mechanism at hand: it does not require the presence of non-collinear magnetic elements and will cause magnetization dynamics in a single ferromagnetic layer \cite{miron}. In addition, it is important to note that whereas different types of domain walls behave in the same way in the absence of spin-orbit interactions, the exact magnetization texture plays a key role when spin-orbit coupling is present due to the coupling between the electron motion and the spin torque. 

In light of the above, several experimental and theoretical works has recently explored the influence of spin-orbit interactions on magnetization dynamics in various magnetic systems \cite{ohe_prl_07, gaididei_pre_10, zhou_pre_08, ryu_jmm_12, janutka_pre_11, moore, miron, santos_pre_11, bijl_arxiv_12}. Since the addition of spin-orbit terms in the equations of motion for the magnetization texture complicates their solution, the large majority of these works have relied on numerical methods to solve the Landau-Lifshitz-Gilbert (LLG) equation. In this paper, we utilize the Lagrangian formalism in order to write down and solve analytically the equations of motion for a domain wall within a collective-coordinate description. The analytical nature of this approach allows us to identify a transparent expression for the domain wall velocity and how it depends on parameters such as the spin-orbit interaction, exchange field, and Gilbert damping of the system. Alternatively, one could have derived this result via the LLG equation, but the present formalism makes it easier to accommodate non-equilibrium spin-density terms and higher order corrections due to the spin-orbit interaction. We show from the analytical expression that the presence of spin-orbit interactions renders the domain wall velocity to be chirality sensitive, in effect depending on whether the wall changes magnetization direction from positive to negative or vice versa along the direction of the current. We provide an estimate for the magnitude of the domain wall velocity using a set of experimentally relevant parameters and show that the velocity behaves qualitatively differently depending on the chirality of the domain wall and displays reentrant behavior. Our finding suggests a way to experimentally distinguish between the non-adiabatic and spin-orbit contribution to the domain-wall speed. 

\begin{figure}
\centering
\resizebox{0.45\textwidth}{!}{
\includegraphics{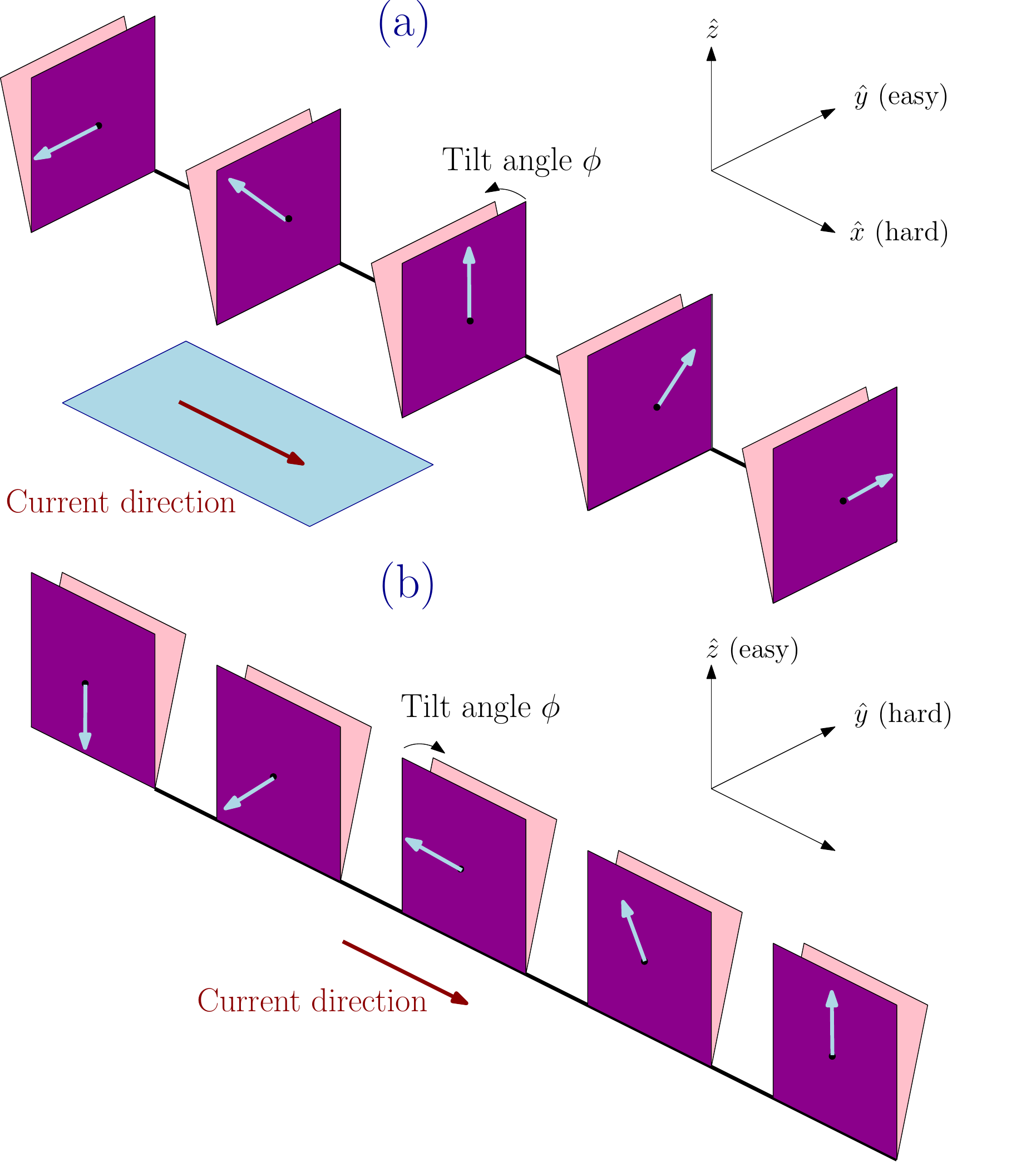}}
\caption{(Color online) Sketch of the magnetization textures considered in this paper. The domain wall nanowire extends along the $x$-axis. (a) Hard axis along current direction $(x)$ and easy axis along $z$-direction. In equilibrium, the magnetization of the domain wall rotates in the $yz$-plane exactly as in Ref. \cite{obata_prb_08} and similarly to Ref. \cite{miron}. Out of equilibrium, i.e. under a current bias, a finite component may be acquired along the $x$-axis as indicated by the tilt angle $\phi$. This geometry is the main focus of this article, relevant for nanowires with perpendicular magnetic anisotropy. (b) Hard axis along $y$-direction and easy axis along $z$-direction. In equilibrium, the magnetization rotates in the $xz$-plane. Out of equilibrium, a finite component may be acquired along the $y$-axis as indicated by the tilt angle $\phi$. }
\label{fig:model} 
\end{figure}

\section{Theory} 

We consider first a ferromagnetic domain wall with an easy (hard) axis of magnetic anisotropy \footnote{The anisotropy energy has the standard form $E_A = -KM_y^2/2 + K_\perp M_x^2/2$, which is assumed to effectively include magnetostatic contributions of the same form as in C. Kittel, Phys. Rev. \textbf{73}, 155 (1948).} along the $y$-axis ($x$-axis). An electric current is injected along the $x$-axis [see Fig. \ref{fig:model}a)], and the inversion symmetry is assumed broken in the $\hat{z}$-direction. This gives rise to a Rashba spin-orbit coupling term, which influences the magnetization dynamics. More precisely, it can be shown \cite{manchon_prb} that the spin-orbit coupling generates an effective magnetic field 
\begin{align}
\boldsymbol{H}_\text{SOC} \propto \alpha_R\hat{\boldsymbol{z}}\times\boldsymbol{j}_e
\end{align}
 where $\alpha_R$ is the Rashba interaction strength and $\boldsymbol{j}_e$ is the current density vector. In order to treat the domain wall as rigid in a collective-coordinate framework, thus making other modes of deformation apart from the tilt angle $\phi$ irrelevant, the easy axis anisotropy energy $K$ is assumed larger than its hard axis equivalent $K_\perp$ \cite{tatara_jpsj_08}, i.e. $|K|\gg|K_\perp|$. As is normally done, we do not take into account the effect of the end-boundaries of the nanowire, assuming thus that the domain-wall center is located sufficiently far away from these during its propagation. 

The starting point is the Lagrangian for such a Bloch domain-wall which was derived in Ref. \cite{obata_prb_08}. For this type of domain wall, the magnetization rotates in the plane perpendicular to the extension of the wire (and current direction), similarly to Ref. \cite{miron} (the only difference from Ref. \cite{miron} is which of the perpendicular axes that is the easy one, in both cases the hard axis is along the current direction). It reads:
\begin{align}
\mathcal{L} &= (\phi\dot{x}-\sin^2\phi) - 2\lambda x J\frac{\Delta}{\mu} - s^y x\dot{\phi}-\phi J\Big(\frac{\Delta}{\mu} + \mathcal{F}(\lambda)\Big)
\end{align}
with the definition
\begin{align}
\mathcal{F}(\lambda) = \frac{\hbar^2\lambda^2}{mL^2}\Big(\frac{1}{\mu} + \frac{2}{\Delta}\Big).
\end{align}
Above, $\phi$ is the tilt angle of the domain wall (see Fig. \ref{fig:model}) while $x$ is the normalized position of the center of the domain wall ($x=X/L$ where $L$ is the wall thickness). The quantities $\lambda$ and $J$ are the Rashba interaction strength and the current density normalized against $mL/\hbar^2$ and $ev_c$, respectively, with $v_c$ being the drift velocity of electrons at the intrinsic threshold current for $\Delta/\mu=1$ without Rashba interactions. Moreover, $\Delta$ is the exchange splitting, $\mu$ is the Fermi energy while $m$ is the electron mass. Finally, $s^y$ represents the constant non-equilibrium spin density induced by the applied electric field generating the current.

\section{Results and Discussion} 

We will model dissipation in this system by a Rayleigh dissipation function of the form \cite{tatara_jpsj_08}
\begin{align}
\mathcal{W} = \frac{\alpha}{2}(\dot{x}^2 + \dot{\phi}^2).
\end{align}
The Lagrange equations are obtained via 
\begin{align}
\frac{\text{d}}{\text{d}t} \frac{\partial \mathcal{L}}{\partial \dot{q}} - \frac{\partial \mathcal{L}}{\partial q} = -\frac{\partial\mathcal{W}}{\partial \dot{q}},\; q\in\{x,\phi\}.
\end{align}
where $t$ is dimensionless time-coordinate normalized against $l/v_c$. Defining for convenience 
\begin{align}
c = \frac{\Delta}{\mu} + \mathcal{F}(\lambda)
\end{align}
we obtain the following Lagrange-equations, which were studied numerically in \cite{obata_prb_08}:
\begin{align}\label{eq:leqs}
\dot{\phi}+\alpha\dot{x} = s^y\dot{\phi} - 2\lambda J\frac{\Delta}{\mu},\;\dot{x}-\alpha\dot{\phi} = -s^y\dot{x} + \sin2\phi + cJ.
\end{align}
The role of the dissipative spin-transfer torque (also known as non-adiabatic or $\beta$-torque) will be addressed later on - we will see that it plays a similar role as the spin-orbit coupling, but with one important difference. By combining the above equations, we are able to eliminate the $\dot{x}$-dependence and obtain a first-order differential equation for the tilt angle: 
\begin{align}
\dot{\phi} = -\frac{\alpha}{1+\frac{\alpha^2}{1+s^y}-s^y}\Big( \frac{\sin2\phi}{1+s^y} + \frac{cJ}{1+s^y} + 2J\frac{\lambda\Delta}{\alpha\mu}\Big)
\end{align}
This equation may be cast into integral form as follows:
\begin{align}
\int\text{d}\phi \frac{\alpha^{-1}(1+s^y)(s^y-1-\alpha^2/(1+s^y))}{a + \sin2\phi} = t.
\end{align}
We define the quantities 
\begin{align}
a &= cJ + \frac{2\lambda\Delta J(1+s^y)}{\alpha\mu},\notag\\
b &= -\alpha-\alpha^{-1}[1-(s^y)^2].
\end{align}
The formal solution of this integral is obtained after some algebraic manipulation:
\begin{align}\label{eq:phi1}
\tan\phi = -a^{-1} + a^{-1}\sqrt{a^2-1}\tan(b^{-1}t\sqrt{a^2-1} + C_0),
\end{align}
where $C_0$ is an integration constant to be determined from the initial conditions. In particular, $\phi(t=0)=0$ and $\phi(t=0)=\pi/2$ yield $C_0=\text{atan}(1/\sqrt{a^2-1})$ and $C_0=\pi/2$, respectively. 
\begin{figure}
\centering
\resizebox{0.49\textwidth}{!}{
\includegraphics{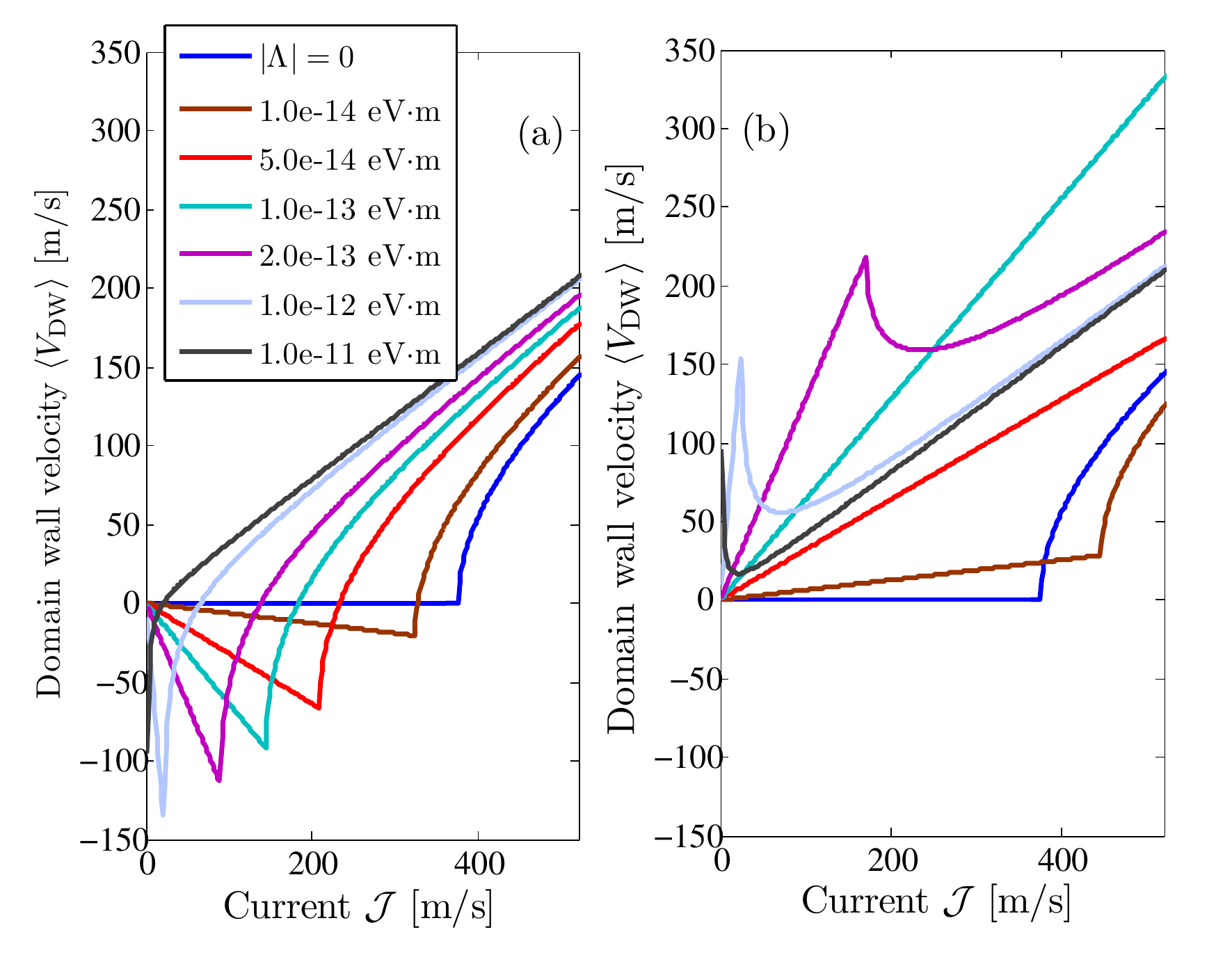}}
\caption{(Color online) Plot of the terminal domain wall velocity using $\alpha=0.005$, $L=75$ nm, $m=0.04m_e$, $\mu = 0.05$ eV, $\Delta=0.02$ eV, and $v_c = 150$ m/s. (a) Positive chirality $\lambda>0$. (b) Negative chirality $\lambda<0$.}
\label{fig:main} 
\end{figure}

Having obtained an explicit expression for the tilt angle, we are now in a position to identify the time-dependence of the domain-wall center $x$, thus also obtaining the domain-wall velocity $\dot{x}$. To do so, we first obtain $\dot{\phi}$ from Eq. (\ref{eq:phi1}) as:
\begin{widetext}
\begin{align}
\dot{\phi} = \frac{(a^2-1)}{ab} \frac{\cos^{-2}(b^{-1}t\sqrt{a^2-1}+C_0)}{1 + [-a^{-1} + a^{-1}\sqrt{a^2-1}\tan(b^{-1}t\sqrt{a^2-1} + C_0)]^2}.
\end{align}
Substituting this into the first Lagrange-equation, one obtains an explicit expression for the domain-wall velocity $v_\text{DW}$
\begin{align}\label{eq:vDW}
v_\text{DW} = -2J\frac{\lambda}{\alpha}\frac{\Delta}{\mu} - (1-s^y)\frac{(a^2-1)}{ab\alpha} \frac{\cos^{-2}(b^{-1}t\sqrt{a^2-1}+C_0)}{1 + [-a^{-1} + a^{-1}\sqrt{a^2-1}\tan(b^{-1}t\sqrt{a^2-1} + C_0)]^2}.
\end{align}
Eqs. (\ref{eq:phi1}) and (\ref{eq:vDW}) determine an exact analytical expression for the time-dependence of the domain wall tilt-angle and velocity, respectively, which we will analyze in more detail below.
\end{widetext}
By integration, Eq. (\ref{eq:vDW}) may be used to identify the time-evolution of the domain-wall center:
\begin{align}
x = -2Jt\frac{\lambda}{\alpha}\frac{\Delta}{\mu} - \frac{(1-s^y)}{\alpha}\int \text{d}t \dot{\phi}.
\end{align}
The integral over $\dot{\phi}$ is evaluated by making use of the formula 
\begin{align}
\int \frac{\text{d}y\cos^{-2}(y+\gamma)}{[1+(\eta+\zeta\tan(x+\gamma))^2]}= \zeta^{-1}\text{atan}[\zeta\tan(x+\gamma)+\eta],
\end{align}
resulting in the following equation describing the instantaneous position of the domain-wall center
\begin{align}\label{eq:xDW}
x-x_0 &= -2Jt\frac{\lambda}{\alpha}\frac{\Delta}{\mu} -\frac{1-s^\gamma}{\alpha}\times\notag\\
&\tan^{-1}[a^{-1}\sqrt{a^2-1}\tan(b^{-1}t\sqrt{a^2-1} + C_0)-a^{-1}],
\end{align}
with $x_0$ being an integration constant related to the initial position of the domain-wall. 

It is worth noting here that there exists a particularly simple solution to the equation set Eq. (\ref{eq:leqs}) in the special case of a constant tilt angle, i.e. a domain-wall that preserves its shape and magnetization direction and thus only propagates. This amounts to setting $\dot{\phi}=0$, which dictates that the tilt angle must be constant: 
\begin{align}
\sin(2\phi_0) = -cJ -2(1+s^\gamma)J\frac{\lambda}{\alpha}\frac{\Delta}{\mu},
\end{align}
and gives for the domain-wall velocity: 
\begin{align}
v_\text{DW} \equiv \dot{x} = -2J\frac{\lambda}{\alpha}\frac{\Delta}{\mu}.
\end{align}
This is identical to the first term in Eq. (\ref{eq:vDW}). Physically, this implies a constant drift velocity of the domain-wall under the application of a current. Interestingly, it is seen that $v_\text{DW}=0$ in the absence of $\lambda$, meaning that the spin-orbit interaction is fully responsible for the domain-wall motion. In a more general scenario where the tilt angle is not restricted to being constant, i.e. allowing for domain-wall deformation, the general expression for the velocity is given by Eq. (\ref{eq:vDW}). As we shall discuss later, this corresponds to the Walker breakdown threshold.

The necessity of spin-orbit coupling to drive the domain-wall motion is a feature pertaining specifically to the Bloch-domain wall with anisotropy axis chosen as described in the beginning of this section [see Fig. \ref{fig:model}a)]. In fact, for different choices of anisotropy directions, spin-orbit coupling mainly contributes as a quantitative correction to the domain-wall velocity without being a prerequisite for its existence. To see this, one may consider instead a domain-wall where the magnetic easy and hard axes lie along the $z$ and $y$-axes, respectively [see Fig. \ref{fig:model}b)]. In this case, the appropriate Lagrangian to consider is \cite{obata_prb_08} 
\begin{align}
\mathcal{L} &= (\phi\dot{x}-\sin^2\phi) +\cos\phi\dot{\phi} - \frac{\pi s^y}{2}(\dot{x}\cos\phi - \sin\phi\dot{\phi})\notag\\
&-s^zx\dot{\phi} - \phi J(\frac{\Delta}{\mu} - \frac{\hbar^2}{2\Delta mL^2}) + \pi\lambda\sin\phi J\Delta/\mu.
\end{align}
Using the same approach as above, one arrives at the following set of integro-differential equations for the time-evolution of the domain-wall center and tilt angle: 
\begin{align}
\dot{\phi} + \frac{\pi s^y}{2}\sin\phi\dot{\phi} + s^z\dot{\phi} = -\alpha\dot{x}
\end{align}
 in addition to 
\begin{align}
\int \text{d}\phi &\alpha^{-1}[\alpha^2 + (1+\frac{\pi s^y}{2}\sin\phi + s^z)^2]/[\pi\lambda\cos\phi J\Delta/\mu \notag\\
&-\sin2\phi - J(\frac{\Delta}{\mu}-\frac{\hbar^2\lambda^2}{2\Delta mL^2})]= t - t_0.
\end{align}
Unlike the situation considered formerly, this equation set may be solved exactly analytically only under simplifying circumstances. For instance, by assuming that the final term in the denominator of the integral equation above dominates and moreover using $s^\gamma\ll1$, $\gamma=\{y,z\}$, one identifies the domain-wall velocity in the strong-current regime $J\gg1$ as 
\begin{align}
v_\text{DW} = -\frac{J}{1+\alpha^2}\Big(\frac{\Delta}{\mu} - \frac{\hbar^2\lambda^2}{2\Delta mL^2}\Big).
\end{align}
As seen, the presence of spin-orbit coupling in this case brings about a minor correction to the final velocity, especially for a strong ferromagnet where $\Delta/\mu$ dominates the expression in parantheses. The domain-wall is driven directly by the current $J$ with a velocity that increases with decreasing dissipation $\alpha$. The above result for the domain wall velocity shows that the spin-orbit coupling has a qualitatively different effect on different domain wall textures.

\begin{widetext}
In the following, we focus on the more interesting case where the spin-orbit coupling influences qualitatively the domain-wall motion [Eqs. (\ref{eq:phi1}) and (\ref{eq:vDW})] and investigate the precise dynamics using the derived analytical expressions. For concreteness, we consider initial conditions such that the tilt angle of the domain-wall at $t=0$ is zero, meaning that we may write:
\begin{align}
\tan\phi &= \frac{1}{a}\Big[\sqrt{a^2-1}\tan\Big(\frac{t\sqrt{a^2-1}}{b} + \text{atan}\Big(\frac{1}{\sqrt{a^2-1}}\Big)\Big) - 1\Big],\notag\\
v_\text{DW} &= -2J\frac{\lambda}{\alpha}\frac{\Delta}{\mu} - \alpha^{-1}(1-s^y)\frac{(a^2-1)}{ab} \times \frac{\cos^{-2}(b^{-1}t\sqrt{a^2-1}+\text{atan}(1/\sqrt{a^2-1}))}{1 + \frac{1}{a^2}[\sqrt{a^2-1}\tan(b^{-1}t\sqrt{a^2-1} + \text{atan}(1/\sqrt{a^2-1}))-1]^2}.
\end{align}
\end{widetext}
In the limit $\alpha\to\infty$, we find that $\phi(t)\to0$ and $v_\text{DW}\to0$ as expected. The analytical expression for $v_\text{DW}$ above reveals that the velocity has non-monotonic behavior as a function of time. In particular, there is a resonance condition $t=t_\text{res}$ at which the velocity increases in magnitude: 
\begin{align}
t_\text{res} = \frac{(n+1/2)\pi b}{\sqrt{a^2-1}} - \frac{b}{a^2-1},
\end{align}
assuming that $a\geq1$. In effect, $v_\text{DW}$ exhibits oscillations which persist even for its terminal behavior $t\to\infty$. It is therefore of interest to establish the average domain wall velocity by averaging over one period $T = \pi b/\sqrt{a^2-1}$: 
\begin{align}
\langle v_\text{DW} \rangle = \frac{1}{T} \int^T_0 \text{d}t v_\text{DW}.
\end{align}
Note that for $a<1$, only the drift-term in $v_\text{DW}$ survives as $t\to\infty$ [since $\cos^2(\pm\i t)\to\infty$ in this limit]. Performing the integral with this in mind, one finds for arbitrary $a$ that 
\begin{align}
\langle v_\text{DW} \rangle = - 2J\frac{\lambda \Delta}{\alpha\mu} - \frac{1-s^y}{\alpha b}\text{sign}\{a\}\mathcal{R}\text{e}\{\sqrt{a^2-1}\},
\end{align}
which written out in terms of the original physical parameters reads
\begin{widetext}
\begin{align}\label{eq:final}
\langle v_\text{DW} \rangle = - 2J\frac{\lambda \Delta}{\alpha\mu} &+ \frac{(1-s^y)\text{sign}\{J\}}{\alpha^2+(1-s^y)}\times\text{sign}\Big\{\frac{2\lambda\Delta (1+s^y)}{\alpha\mu} + \frac{\Delta}{\mu} + \mathcal{F}(\lambda)\Big\}\times\mathcal{R}\text{e}\Bigg\{\sqrt{\Big[ \frac{2\lambda\Delta (1+s^y)}{\alpha\mu} + \frac{\Delta}{\mu} + \mathcal{F}(\lambda)\Big]^2J^2 - 1} \Bigg\}
\end{align}\end{widetext}
Eq. (\ref{eq:final}) is the main result of this work and constitutes a generally valid analytical expression for the terminal domain wall velocity taking into account both Rashba spin-orbit coupling, the non-equilibrium spin density, and Gilbert damping. For consistency, we have verified that a fully numerical solution of the equations of motion give identical results as the above analytical expression for the domain wall velocity.

We will proceed to analyze this velocity quantitatively for a realistic set of parameters and investigate how it depends in particular on the applied current and the magnitude of the spin-orbit coupling. Before doing so, one should note that it follows from Eq. (\ref{eq:final}) that there exists both a threshold current $J_c$ for which the second term in Eq. (\ref{eq:final}) is non-zero:
\begin{align}\label{eq:walker}
|J_c| &= \Big| \frac{2\lambda\Delta (1+s^y)}{\alpha\mu} + \Big(\frac{\Delta}{\mu} + \frac{\hbar^2\lambda^2}{mL^2\mu} + \frac{2\hbar^2\lambda^2}{mL^2\Delta}\Big) \Big|^{-1}.
\end{align}
In the limiting case of zero spin-orbit interaction, one obtains $|J_c| = (\Delta/\mu)^{-1}$ in agreement with previous studies. The presence of spin-orbit interaction is seen from the analytical expression of $|J_c|$ to reduce the threshold current monotonically with increasing $\lambda$, consistently with the numerical study in Ref. \cite{obata_prb_08}. This monotonic behavior appears also when tuning the chemical potential $\mu$: increasing $\mu$ lowers the polarization and increases the threshold current. 

Eq. (\ref{eq:walker}) is in fact the Walker threshold value which separates the regimes of domain wall motion with a fixed profile, i.e. $\dot{\phi}=0$ and the regime with a domain wall rotating its spatial profile as time increases, i.e. $\dot{\phi}\neq0$. To see this, one may revert to the original equations of motion in Eq. (\ref{eq:leqs}). There exists a tilt angle $\phi$ which satisfies $\dot{\phi}=0$ if the following equation is satisfied:
\begin{align}
\sin2\phi = -\Big(\frac{2\lambda\Delta (1+s^y)}{\alpha\mu}J + cJ\Big).
\end{align}
Since the left-hand side varies between $\pm1$, one may find a solution if the following equation holds:
\begin{align}
|J|\Big|\frac{2\lambda\Delta (1+s^y)}{\alpha\mu} + c\Big| < 1.
\end{align}
which is completely equivalent to Eq. (\ref{eq:walker}) after rewriting. For larger currents $J$, there exists no time-independent solution $\phi$ and domain wall distortion $\dot{\phi}$ is now inevitable past the Walker breakdown. 

It has previously been suggested that the spin-orbit interaction and non-adiabatic spin-torque influence the magnetization dynamics in the same manner, since the latter may be included by substituting $\lambda \to \beta + \lambda$ \cite{obata_prb_08}. However, it was noted in Ref. \cite{ryu_jmm_12} that the chirality of the domain wall determines the effective sign of the spin-orbit coupling $\lambda$ in the equations of motion. Formally, this corresponds to taking the domain wall profile represented with polar angles $\phi$ and $\theta$ as $\boldsymbol{M} = M_0(-\sin\theta\sin\phi\hat{\boldsymbol{x}} + \cos\theta\hat{\boldsymbol{y}} + \sin\theta\cos\phi\hat{\boldsymbol{z}})$ and performing the transformations $\phi \to (-\phi)$ and $\cos\theta\to(-\cos\theta)$. The parameter $\phi=\phi(t)$ is the time-dependent tilt angle, whereas $\theta$ is defined by 
\begin{align}
\sin\theta = \text{sech}[(\tilde{x}-X(t))/L_\text{DW}],
\end{align}
 where $\tilde{x}$ is the position along the magnetic wire, $X(t)$ is the time-dependent center position of the domain-wall, whereas $L_\text{DW}$ is the domain wall width. This suggests that the role of spin-orbit coupling is chirality-sensitive and in this regard differs qualitatively from non-adiabaticity. Below, we will investigate this effect and show that the chirality indeed gives rise to highly different behavior for the domain wall velocity. The chirality is changed in our analytical expression Eq. (\ref{eq:final}) by letting $\lambda \to (-\lambda)$. We note that our conclusions remain unchanged even when including the $\beta$-term, as long as it is small (which is typically the case).

In the remainder of this paper, we fix the following parameters: $\alpha=0.005$, $L=75$ nm, $m=0.04m_e$, $\mu = 0.05$ eV, $\Delta=0.02$ eV, and $v_c = 150$ m/s to model an experimentally realistic semiconductor system \cite{yamanouchi_nature_04}, where $m_e$ is the bare electron mass. We restrict our attention to a scenario where the electron-spin density satisfies $s^y\ll1$. In Eq. (\ref{eq:final}), note that $J$ and $\lambda$ are normalized quantities. To make a quantitative estimate for the domain wall velocity, we define the non-normalized current $\mathcal{J} = Jv_c$ and spin-orbit interaction $\Lambda = \hbar\lambda/(mL)$ which have units m/s and eV$\cdot$m, respectively. Similarly, we restore the dimension of the terminal domain wall velocity by defining $\langle V_\text{DW} \rangle = v_c\langle v_\text{DW} \rangle$ with units m/s. 

We show in Fig. \ref{fig:main} a plot of the terminal domain wall velocity as a function of the applied current for several values of the spin-orbit interactions. To illustrate the effect of the chirality, we plot in (a) the case $\Lambda>0$ while in (b) $\Lambda<0$. The latter results are consistent with the numerical study of Ref. \cite{obata_prb_08}, and shows that the spin-orbit interaction can greatly enhance the domain wall velocity at low currents up to the Walker breakdown. In the former case, however, the domain wall velocity behaves differently when changing the current $\mathcal{J}$. The spin torque induced by the effective Rashba-field is now directed opposite to the conventional current-driven spin torque and there is a competition between the two. For non-zero $\Lambda$, the domain wall velocity starts by moving in the direction \textit{opposite} to the current, whereas it eventually changes sign with increasing $\mathcal{J}$ (above the threshold current). Thus, experimentally observing a sign-reversal of the domain-wall velocity with applied current would be an indication of precisely this chirality sensitive spin-orbit coupling effect. We note that this sign-reversal of the wall velocity is different from the one predicted in \cite{ryu_jmm_12}, which originated from a Slonczewski-like spin-orbit torque proportional to $\beta$ and when considering a different type of domain wall profile. In our case, the field-like spin-orbit torque is sufficient to cause the velocity-reversal. 

\section{Summary} In conclusion, we have used the Lagrangian formalism to derive an exact analytical expression for the domain wall velocity in a spin-orbit coupled ferromagnet. We have shown that a chirality-sensitive domain wall velocity appears in this system, which qualitatively differs from the role of the non-adiabatic spin-transfer torque stemming from a spatial mistracking between the conduction electrons and the local magnetization. A candidate system for the observation of this effect would be hybrid structure comprised of a thin ferromagnetic wire in contact with a heavy metal and an oxide, which would break structural inversion symmetry and this provide a gradient in the electric potential which generates the Rashba spin-orbit coupling. This type of structure has indeed recently been experimentally considered in Refs. \cite{moore, miron} (Pt/Co/AlOx) whereas similar systems were studied in Refs. \cite{ikeda, shiota, wang} (Ta/CoFeB/MgO).\\
\text{ }\\

\acknowledgments

The author would like to thank G. Tatara and J. Ryu for a clarifying correspondence.

\end{document}